\documentclass[11pt]{article}

\usepackage[utf8]{inputenc}
\usepackage[T1]{fontenc}
\usepackage{mathptmx}
\usepackage[a4paper,margin=2.5cm]{geometry}
\usepackage[numbers,sort&compress]{natbib}

\usepackage{xcolor}
\usepackage{hyperref}
\definecolor{darkolivegreen}{rgb}{0.33, 0.42, 0.18}
\definecolor{celestialblue}{rgb}{0.29, 0.59, 0.82}

\hypersetup{colorlinks,
            linkcolor=darkolivegreen,
            citecolor=darkolivegreen,
            urlcolor=darkolivegreen,
            pdftitle={Dynamic Reduced-Order Data Assimilation from Sparse Velocity Measurements},
            pdfauthor={Mauricio Portilla, Felipe Galarce, Ernesto Castillo, and Benjamin Herrmann}}
\usepackage{amsmath, amsthm, amssymb, bm}
\usepackage{thmtools, thm-restate}
\usepackage{dsfont}
\usepackage{mathtools}
\usepackage{graphicx}   
\usepackage{color}
\usepackage{setspace}
\usepackage{algpseudocode}

\newenvironment{algorithm}{\begin{table}}{\end{table}}

\theoremstyle{definition}

\newcommand{\cA}{\ensuremath{\mathcal{A}}}

\newcommand{\cH}{\ensuremath{\mathcal{H}}}

\newcommand{\cM}{\ensuremath{\mathcal{M}}}
\newcommand{\cN}{\ensuremath{\mathcal{N}}}

\newcommand{\cP}{\ensuremath{\mathcal{P}}}

\newcommand{\bA}{\ensuremath{\mathbb{A}}}

\newcommand{\bC}{\ensuremath{\mathbb{C}}}

\newcommand{\bH}{\ensuremath{\mathbb{H}}}

\newcommand{\bR}{\ensuremath{\mathbb{R}}}

\def\[{\left[}
\def\]{\right]}
\def\<{\langle}
\def\>{\rangle}
\def\({\left(}
\def\){\right)}
\def\[{\left [}
\def\]{\right]}
\def\({\left(}
\def\){\right)}

\newcommand{\norm}[1]{\Vert #1 \Vert}

\newcommand{\bu}{\ensuremath{\boldsymbol{u}}}
\providecommand{\press}{\ensuremath{p}}

\def\xo{\textbf{X}_0}
\def\xi{\textbf{X}_1}
\def\riT{\textbf{W}^T}
\def\ri{\textbf{W}}
\def\podu{\textbf{U}}
\def\pods{\Sigma}
\def\podw{\mathbb{W}}
\def\koop{\textbf{A}}
\def\pkoop{\tilde{\textbf{A}}}
\def\dmdmodes{\Psi}

\def\dmdb{\textbf{b}}
\def\pdmdmodes{\tilde{\textbf{V}}}

\def\crossg{\textbf{G}}
\def\nbmodes{n}

\def\nbsnaps{N_s}
\def\nbdofs{\mathcal{N}}
\def\nbparams{N_p}

\usepackage{graphicx}
\usepackage{dcolumn}
\usepackage{bm}
\usepackage{etoolbox}
\begin{document}

\title{Dynamic Reduced-Order Data Assimilation from Sparse Velocity Measurements}

\author{Mauricio Portilla$^{1}$, Felipe Galarce$^{1}$\footnote{Corresponding author: \url{felipe.galarce@pucv.cl}}, Ernesto Castillo$^{2,3}$, and Benjamin Herrmann$^{4,5}$\\[0.5em]
\small $^{1}$School of Civil Engineering, Pontificia Universidad Católica de Valparaíso, Valparaíso, Chile\\
\small $^{2}$Departamento de Ingeniería Mecánica, Universidad de Santiago de Chile, Santiago, Chile\\
\small $^{3}$Computational Heat and Fluid Flow Lab, Universidad de Santiago de Chile, Santiago, Chile\\
\small $^{4}$Department of Mechanical and Metallurgical Engineering, Pontificia Universidad Católica de Chile, Santiago, Chile\\
\small $^{5}$Department of Hydraulic and Environmental Engineering, Pontificia Universidad Católica de Chile, Santiago, Chile\\[0.5em]
\small Corresponding author: \texttt{felipe.galarce@pucv.cl}}

\date{\today}

\maketitle

\begin{abstract}
We present a novel reduced-order data assimilation framework, termed Reduced-Order Dynamical Assimilation (RODAS), for reconstructing high-resolution, time-resolved flow fields from sparse velocity measurements. The method combines low-dimensional experimental observations with a physics-based parametric reduced-order model, enabling both spatial extrapolation beyond the measurement region and temporal super-resolution. The approach first identifies the dominant dynamics from sparse measurements through Dynamic Mode Decomposition (DMD), and subsequently reconstructs the corresponding full-order flow evolution by projecting the identified dynamics onto a parametric Proper Orthogonal Decomposition (POD) manifold generated from high-fidelity numerical simulations. Unlike conventional reduced-order data assimilation methods that estimate independent snapshots or treat time as an additional parameter, RODAS reconstructs an entire dynamical trajectory in a single inference step while naturally incorporating parametric variability. We assess the proposed methodology on vortex shedding behind a circular cylinder for both Newtonian and non-Newtonian (Carreau–Yasuda) fluids. Numerical experiments demonstrate accurate reconstruction of high-resolution velocity fields from localized, low-resolution measurements, achieving sub-percent reconstruction errors with sufficiently rich reduced bases, robust performance under severe temporal undersampling, and accurate prediction of engineering quantities of interest such as the drag coefficient. These results demonstrate that RODAS provides an efficient framework for real-time, physics-informed reconstruction of unsteady flows from sparse experimental data. 
\end{abstract}

\noindent\textbf{Keywords:} data assimilation, reduced-order modeling, dynamic mode decomposition, sparse measurements, non-Newtonian flow

\section{\label{sec:level1}Introduction}

Scientific machine learning (SML) has emerged as a relevant tool that leverages advances in artificial intelligence, applied statistics, and computational mechanics \cite{Brunton_Kutz_2019} to longstanding problems in science and engineering. To name a few: civil engineering \cite{TORZONI2024116584}, metallurgy \cite{beckermann2020}, the food industry \cite{szpicer2023, GALARCE2025110374}, meteorology \cite{marco_2020}, and biomechanics \cite{quarteroni24_cardio, Morris18, MINSAL2021}. In many applications, the need to repeatedly evaluate complex high dimensional models has led to the development of reduced-order models (ROM), \cite{Brunton2024, hesthaven2016, benner2017model, Wilcox2010}, as a tool to capture essential characteristics of the full system with lower complexity and user-controlled error. 

This work is not only focused on ROMs, but also on the use of a reduced-basis within the task of estimating a field state from experimental measurements (or inverse problem), rather than creating a map from the parameter space to the solutions (forward problem). The integration of ROMs and experimental measurements can be formulated as a data assimilation problem \cite{asch2016_dataAss}, combining both physical models and measurements, to provide accurate and fast predictions on a system state or parameters \cite{lahoz2010data, lombardi_2023_review, Lemus_herrmann_2025}. The goal of this paper is to efficiently assimilate under-sampled and eventually noisy velocity data in moderate Reynolds non-Newtonian flows. Current approaches using ROMs typically do not leverage the time-marching nature of the phenomena, thus yielding methods more suitable to stationary set-ups \cite{GLM2022, HAIK2023115868}. In contrast with these methods, we do not treat time as an additional parameter of the solution manifold. Instead, we discover the full dynamical trajectory in a single assimilation task, in a space-time manner.

Our approach is based on learning the trajectory of a dynamical system from experimental data so that we can next map it to a high-dimensional parametric manifold. From a methodological standpoint, we achieve this by means of a least-squares fit between the measurements and linear dynamical modes. The challenge is thus twofold: First, we try to recover a high-fidelity state from partial measurements, yielding an ill-posed problem \cite{BCDDPW2017} which is cured using the physical equations as a prior background \cite{galarce2023bias}. Second, we deal with time under-sampled data, and we rely on a linearized modal operator coming from both the dynamic mode decomposition of the sensor data, plus a parametric fit in a high-dimensional space. 

\subsection{Model order reduction}

The growing availability of high-fidelity data from numerical simulations and experimental measurements of fluid flows has spurred the development and widespread adoption of various data-driven modal decompositions~\cite{taira2017aiaa,taira2019aiaa}. Among the earliest and most commonly used in fluid dynamics is the proper orthogonal decomposition (POD), originating from the application of stochastic methods to turbulence by Lumley~\cite{lumleybook,berkooz1993arfm}. Its enduring popularity stems from its straightforward implementation and the fact that the resulting modes are orthogonal and hierarchically ordered according to the amount of sustained energy they represent in the flow. Some other ROM strategies include the randomized singular value decomposition \cite{Halko2011}, data-driven ROMs \cite{GGLM2021}, the generalized empirical interpolation method (GEIM) \cite{MMT2016, BARRAULT2004667}, among others: \cite{Wilcox2024, RHP2007, peherstorfer2015, CD2015acta}.

The dynamic mode decomposition (DMD) is another important data-driven technique that emerged within the fluid dynamics community~\cite{schmid2010jfm,rowley2009jfm}. Over more than a decade, DMD has become a standard tool for analyzing fluid flows, as it simultaneously identifies coherent flow structures and constructs a linear model for their dynamics from time-resolved measurements~\cite{schmid2010jfm,rowley2009jfm,schmid2022arfm}. During this period, significant efforts have been made to extend the applicability of DMD to nonlinear dynamical systems by leveraging its deep relationship with Koopman theory~\cite{rowley2009jfm,mezic2013arfm,brunton2022siamrev}. These efforts have enabled the use of non-sequential data~\cite{tu2014jcd}, the promotion of sparsity~\cite{jovanovic2014pof}, the handling of streaming datasets~\cite{hemati2014pof}, the enhancement of accuracy through nonlinear observables~\cite{williams2015jns}, the inclusion of control effects~\cite{proctor2016jads}, the improvement of robustness via nonlinear optimization~\cite{askham2018jads}, the modeling of traveling and standing waves with higher-order SVD~\cite{leclainche2018jns}, and the scalability to large datasets using randomized linear algebra~\cite{erichson2019jads}, among other developments~\cite{dmdbook,schmid2022arfm}.

A recent advancement, physics-informed DMD (piDMD), integrates known physical principles—such as symmetries, invariances, and conservation laws—directly into the DMD framework~\cite{baddoo2023prsa}. This is achieved by recasting the underlying optimization as a Procrustes problem and constraining the admissible models to a matrix manifold that preserves the desired physical structure~\cite{baddoo2023prsa}. Consequently, piDMD models tend to exhibit reduced overfitting, require less training data, and are often more computationally efficient to construct than traditional DMD models~\cite{baddoo2023prsa}.

\subsection{State estimation and manifold learning}

The problem of estimating a field or a set of parameters from sensor data is widely studied \cite{lahoz2010data, romor2026}. The contemporary formulation of this problem is as follows: given an element in an \textsl{observation space}, find the best state or full state trajectory in a \textsl{manifold of solutions} stemming from the governing equations. This setup is a so-called \textsl{manifold learning} task, and its study has gained attention over the last decade \cite{romor2023_nonLinearManifoldLearning}. We will adopt this framework to describe our approach in the following sections. 

The literature offers several alternatives to deal with this manifold learning endeavor, specially when it concerns the estimation from the parameter space as a starting point (a \textsl{Forward Problem}). This is the case of several machine learning pipelines, like convolutional auto-encoders \cite{LEE2020108973}, physics-informed neural networks (PINNs) \cite{ROJAS2024116904, RAISSI2019686}, or U-Net architectures for the prediction of unsteady flows \cite{YANG2026131843}. When integrated with experimental data, the literature is a bit more scarce, yet some works can be found on this matter, such as using PINNs for the estimation of rheology-dependent parameters \cite{sierpe2025estimationhemodynamicparametersphysics}, estimating the mechanical response in human organs using ROM-based variational approaches \cite{Karabelas2026_cardiac}, or deep-learning enhanced reduced order models \cite{BRIVIO2025117989}.

Similarly, there are classical methods dealing with data assimilation problems. Two famous examples are 4D-var \cite{marco_2020,moore_2011} and Kalman filters \cite{evensen2009data, bertoglio_fsi_assim}, which enable sequential estimation in real time, or a full integration of time-dependent measurements in a single optimization problem. 

We next describe the mathematical framework we adopt to understand our manifold learning problem, including a linear representation to build the observation space, the building of a parametric proper orthogonal decomposition basis, and the DMD inference from sparse measurements.

\section{Inverse problems with parametric model reduction}

This work focuses on the task of recovering a state $\bu$ from a sparse set of experimental measurements, assuming they are low-resolution in both, time and space. The state $\bu$ is assumed to belong to a Banach or Hilbert space $V$, defined over a spatial domain $\Omega \in \bR^d$ ($d = 2,3$), with inner product $\left(\cdot,\cdot\right)$ and induced norm $\lVert \cdot \rVert = \sqrt{\left(\cdot,\cdot\right)}$. 

As it is usual in modern scientific machine learning, we also assume that our search is guided by a physics-informed solution manifold \cite{romor2023_nonLinearManifoldLearning}, given by a partial differential equation $\cP$:
\begin{equation}
\label{eq:manifold}
\cM = \{ \bu \in V; \, \cP(\bu, \theta) = 0; \, \theta \in \Theta\},
\end{equation}
where $\cP$ stands for the prior physical knowledge of the dynamical system we aim to recover, and $\Theta$ is the parameter space. The reader might have noticed that, by introducing this continuous setting, we can describe the method without choosing a priori any sort of discretization strategy for the governing equations. This way, we will see methods such as the finite element method (FEM), or the finite volume method (FVM), as a sampling strategy of the discrete version of $\cM$. 

The experimental measurements are assumed linear \cite{BCDDPW2017, HAIK2023115868}. That is to say, there exist linear functionals $\ell_i(\bu(t)): V \rightarrow \bR$, such that:
\begin{equation}
\ell_i(\bu(t)) = (\bu(t), \lambda_i),
\label{eq:riesz}
\end{equation}
where $i=1,\ldots,m$, and $m$ is the number of spatial sensors, which are assumed to be fixed in space, and $\lambda_i$ the corresponding Riesz representers, that span the observation space, denoted $W_m$ \cite{GGLM2021}. The evolution of the measurements with time is modeled with an operator $\cH$, so that:
\begin{equation}
\frac{d\ell_i(t)}{dt} = \cH(\ell_i(t)).
\label{eq:measurements_evolution}
\end{equation}

We describe next a methodological framework to perform state estimation from the measurements given physical and experimental information, highlighting the following:
\begin{itemize}

\item The core of the method stems from a strategy that approximates, from the measurements only, a linear operator for the dynamics of the system, thus yielding a flexible reconstruction which does not require any a-priori assumption on the parameter configuration the measurements corresponds to. 


\item The measurements will be sparse in time and space, so we will be able to extrapolate our estimations from localized, low-resolution observations, to high-fidelity fields with continuous time evaluation.
\end{itemize}

Our idea comes from comparing this evolution problem with one related to the full-order physical system, using another operator $\cA : V \rightarrow V$, such that:
\begin{equation}
\frac{d\bu(t)}{dt} = \cA(\bu(t)),
\end{equation}
and assuming that the spectral content of $\cA$ is similar to that of $\cH$ in Eq.~(\ref{eq:measurements_evolution}), in the sense that the relevant eigenvalues of $\cH$ are also present in the observed operator spectral decomposition. This is a reasonable hypothesis on the observability of the phenomena. That is to say, we assume that the sensors have enough quality to capture dominant frequencies of the full phenomena.

We propose a Reduced-Order Dynamical ASsimilation (RODAS) strategy, which entangles some properties of $\cH$ with those of $\cA$, enabling spatio-temporal super-resolution of the measurements time-series. 

In particular, we work with linearized operators associated with both evolution problems, that we will denote by $\bH$ and $\bA$, for the measurements and the full state continuous-time propagators, respectively. For practical reasons, as we aim to apply our methodology in engineering problems, we will sometimes drop the infinite-dimensional formulation to work with real-valued vector states $\bu \in \bR^\cN$. We will not change the notation for the finite dimensional ambient space, so we will refer to $V$, when no ambiguity arises, as either the infinite-dimensional space or the finite dimensional approximation (typically arising, for example, in finite element solvers). We next assume that the frequency content of $\bA$, which we can approximate using a dynamic mode decomposition of the low-resolution data, is similar to that of the sub-sampled data evolution operator $\bH$.

We address two main challenges in this work. On the one hand, we have already discussed the time sub-sampling of the experimental data, which could lead to unobservable modes, yielding very ill-posed inverse problems. On the other hand, we also extrapolate the experimental measurements beyond the sensor acquisition region, which is usually a sub-domain $\Omega^{\text{measures}} \subset \Omega$, much smaller than $\Omega$ and where the measurements are defined with a lower spatial resolution compared to the one typically associated with the numerical solution of a PDE. 

This sort of extrapolation problem is tackled using information from the solution manifold $\cM$, which we will assume admits a linear model order reduction with a sub-space $V_n \subset V$ such that $\max_{u \in V} \norm{u - \Pi_{V_n}u}_V \leq \epsilon_n$, i.e., with some control of the error $\epsilon_n$ induced by the model reduction. Hereafter, $\Pi_X$ denotes an orthogonal projection onto some closed sub-space $X$.  This assumption is justified theoretically by the fast-decaying Kolmogorov $n-$width of the solution manifold, which is the case of several fluid dynamics applications, including moderate Reynolds Navier-Stokes equations \cite{GGLM2021}.

Let $\rho_1,\ldots,\rho_n$, be a basis for $V_n$, and let us consider the eigen-decomposition $\bA \psi_i = \psi_i \Lambda_i$, with $\Lambda \in \bC^{\cN \times \cN}$. Our claim is that we can approximate this decomposition as follows:


\begin{enumerate}
    \item Equal frequencies: we impose that the subdomain modes are observations of the full domain modes, which implies that $\Lambda_{meas} = \Lambda$
    \item Least-squares extrapolation of the modes using parametric ROM: we do assume that each eigen-vector of $\bA$ can be re-cast as a linear combination of the parametric basis of $V_n$, meaning in practice that we solve for full domain modes $\Psi$.
\end{enumerate}

Since we rely on a dynamic mode decomposition of both the measurements and the full system trajectory, we recall the classical DMD formulation, so we can next introduce the least-squares fit to do data assimilation.

We remark that, while the algorithm is described for generalized governing equations in \eqref{eq:manifold}, we will pay particular attention to the case of fluid mechanics, in which the gold-standard dynamics is the incompressible Navier-Stokes equations. Throughout this work, we regard the nonlinear evolution problem as solving for the fields $\bu: \Omega \rightarrow \bR^d$ and $p: \Omega \rightarrow \bR$ such that:
\begin{align}
    \label{eq:momentum}
    \rho \partial_t \bu + \rho \bu \cdot \nabla \bu + \nabla \press - \nabla \cdot \tau = 0, \quad \text{ in } \Omega,\\
    \label{eq:mass}
    \nabla \cdot \bu = 0, \quad \text{ in } \Omega,
\end{align}
where $\rho$ stands for the fluid density and $\tau$ stands for the shear stress tensor. In our case, we will deal in our numerical examples with two prototypical cases:
\begin{itemize}
    \item Newtonian model: The shear rate depends linearly on the strain rate, using $\tau = \mu  \nabla^s \bu $, with $  \nabla^s = 1/2(\nabla \bu + \nabla^T \bu)$, the symmetrized velocity gradient. This implies a classical Newtonian rheology, which along the incompressibility assumption, leads to a Laplacian term in the governing equations \eqref{eq:momentum}, \eqref{eq:mass}, i.e. $\nabla \cdot \tau = \mu \Delta \bu$, where $\mu$ stands for a constant viscosity. 
    \item Carreau-Yasuda model: An effective viscosity $\eta$ is instead computed in the following non-linear form:
   \begin{align}
     \eta(\dot{\gamma}) = \nu_{\mathrm{min}}  + (\nu_{\mathrm{max}} - \nu_{\mathrm{min}})\left[1 + \left(\lambda\dot{\gamma}\right)^a\right]^{\frac{n-1}{a}},
\end{align}
     where $a > 0$, $n\in [0,1)$, $\lambda > 0$ and $\nu_{\mathrm{max}} > \nu_{\mathrm{min}} > 0$ are empirical constants. The viscous force is then introduced as $\nabla \cdot \tau = \nabla \cdot (\eta(\dot{\gamma}) \nabla^s \bu)$ in equation \eqref{eq:momentum}.
\end{itemize}
We will solve \eqref{eq:momentum}, \eqref{eq:mass} for a large sample of the parameter space, building physically coherent datasets for model order reduction. The boundary conditions to close the system are discussed in section \ref{sec:numerical_examples}.

\subsection{Dynamic mode decomposition}

Let $\{ l_1, \ldots, l_{N_s} \}$ be a single trajectory of $\nbsnaps$ snapshots coming from experimental measurements, each with $m$ degrees of freedom. We can regard these snapshots as time marching solutions of the governing equations (via the finite element or finite volume methods, for instance) for a single parameter configuration. We can gather the solution in offset matrices $\xo$ and $\xi$ in $\mathbb{R}^{m \times \nbsnaps - 1}$ such that:

    \begin{equation}
        \xo=
        \begin{bmatrix}
          \vert &  & \vert \\
          u_1   & ...  & u_{N_s-1} \\
          \vert &  & \vert
        \end{bmatrix}
          \; \text{and, }
      \xi=
    \begin{bmatrix}
      \vert &  & \vert \\
      u_2   & ...  & u_{N_s} \\
      \vert &  & \vert
    \end{bmatrix}
    \; .
    \end{equation}

    At the core of DMD is the assumption that the observed dynamics can be well approximated by a linear evolution operator, even when the underlying system is nonlinear. In a discrete setting, this amounts to finding a matrix $\koop \in \bR^{m \times m}$ that advances the observed state from one snapshot to the next:
    \begin{equation}
        \label{eq:argminA}
        \koop = \mathop{\arg \min}\limits_{\text{rank}(\boldsymbol{B})=r} \lVert \xi - \boldsymbol{B} \xo \rVert_2 \;,
\end{equation}
whose solution is obtained using the Moore-Penrose pseudo-inverse, i.e.,  $\koop^{*}=\xi\xo^{\dagger}$.  Next, using a singular value decomposition (SVD) in $\xo=\podu\pods\podw^T$, where $\podu$ and $\podw$ contain the left and right singular vectors, and $\pods$ is a diagonal matrix with singular values, and keeping the first $r < m$ dominant modes we define the truncated matrices $\podu_r$, $\pods_r$ and $\podw_r$. In practice, one approximates: $\koop = \xi \podw_r \Sigma_r^{-1} \podu_r^T$. In fact, for the sake of computational cost, we project $\koop$ onto the column space of $\podu_r$, we obtain the reduced operator $\pkoop \in \mathbb{R}^{r \times r}$, essentially a pull-back to the full evolution in $\koop$: $\podu_r^T \koop \podu_r = \podu_r \xi \podw_r \Sigma_r^{-1}$. We consider the decomposition $\pkoop \pdmdmodes = \pdmdmodes \Lambda$, where $\Lambda \in \mathbb{C}^{r \times r}$ contains the DMD eigenvalues and $\pdmdmodes \in \mathbb{C}^{r \times r}$ the corresponding modes. 

As $\pkoop$ acts on the subspace spanned by $\podu_r$, the full DMD modes can be approximated as a linear combination of the modes in $\podu_r$, yielding $\Psi \approx \podu_r \tilde{\textbf{V}}$. We take inspiration from this very fact to build our assimilation strategy, by replacing the basis in $\podu_r$ by a global parametric and extended basis stemming from a database of numerical simulations. Our idea is then to design an appropriate algorithm to find an element in this new extended sub-space, finding its coordinates via assimilation of a measurement time-series.

\subsection{Parametric DMD assimilation}
\label{sec:rodas}

We can now write down our assimilation strategy, the Reduced-Order Data Assimilation (RODAS) approach, which builds a least-squares fit from the experimental dynamic mode decomposition $\Psi^{meas}$ and the full domain DMD $\Psi$. The observation process, i.e., the projection onto the space $W_m$ is written in finite-dimension using the Riesz-representers from \eqref{eq:riesz}, placing them in the columns of a matrix $\ri \in \bR^{\cN \times m}$. 

Assuming that performing DMD on a subdomain yields approximate observations of the full-domain modes with comparable eigenvalues, we can define the following optimization problem for the i-th mode $\dmdmodes_i$:

\begin{equation}
     \Psi_{i}^{*} = \mathop{\arg \min}\limits_{\Psi_{i} \in \mathbb{C}^{\nbdofs}} \lVert \riT \Psi_i - \Psi_{i}^{meas} \rVert_2,
\end{equation}
where $\dmdmodes_i^{meas}$ denotes the i-th mode obtained from the measurement domain, i.e., the projection of the unknown full-state mode $\dmdmodes_i$. The problem is ill-posed since $\ri \riT$ is generally non-invertible, implying that multiple full-state modes could produce identical measurement projections.

To regularize the problem, we introduce prior information from simulations in $\cM$ with varying parameters $\theta_i$, $ i=1,...,p$. Let $\Phi\in\mathbb{R}^{\nbdofs\times\nbmodes}$ be a unitary matrix whose columns store the parametric POD modes obtained from these training simulations.

We assume that the full-state dynamics can be approximately represented in the subspace spanned by $\Phi$. The full-state DMD modes can then be recovered by finding the linear combination of the POD basis vectors that best reproduces the observed subdomain modes:

\begin{equation}
    \label{eq:minprob}
    \tilde{\textbf{V}}_p^{*} = \mathop{\arg \min}\limits_{\tilde{\textbf{V}}_p \in \mathbb{C}^{{\nbmodes} \times {\nbmodes}}} \lVert \textbf{W}^T \Phi \tilde{\textbf{V}}_p - \Psi^{meas}\rVert_2.
\end{equation}

Here, each column of $\tilde{\textbf{V}}_p$ represents the coefficients of one reconstructed DMD mode in the parametric reduced order column space of $\Phi$.

Defining $\crossg=\riT\Phi$, the least-squares solution of the previous problem satisfies the equation:
\begin{equation}
    \crossg^T\crossg\pdmdmodes_p = \crossg^T\Psi^{meas}\;,
\end{equation}
which can be solved independently for each column of $\pdmdmodes_p$. The reconstructed full-domain DMD modes can be approximated with

\begin{equation}
\label{eq:parametricPOD}
    \Psi \approx \Phi\pdmdmodes_p \; .
\end{equation}

To compute the initial amplitudes of the reconstructed modes, we approximate the first full-state snapshot $\bu_0$.
Given that only its projection $\riT \bu_0$ is available from measurements, we solve:
\begin{equation}
    \textbf{c}^* = \mathop{\arg \min}\limits_{\textbf{c} \in \mathbb{R}^{\mathcal{N}}} \lVert \riT \left( \Phi \textbf{c} - \bu_0 \right) \rVert^2,
\end{equation}
whose normal equations are:
\begin{equation}
    (\crossg^T\crossg)\textbf{c}=\crossg^T\riT \bu_0.
\end{equation}


Let $\dmdb \in \mathbb{C}^{r}$ denote a vector that contains the initial DMD mode amplitudes such that $\bu_0 \approx\dmdmodes\dmdb$. The reconstructed state at any time $t$ is then given by:
\begin{equation}
    \label{eq:dmdrec}
    x(t) =\sum_{k=1}^{r} \psi_k \text{exp}(\omega_k t) b_k,
\end{equation}
where $\omega_k = \text{log}(\lambda_k)/\Delta t$ and $\lambda_k$ is the k-th DMD eigenvalue. A compact algorithmic view of the data assimilation procedure is depicted in Tab.~\ref{alg:pdmda}.

\begin{algorithm}
\caption{Reduced-Order Dynamical Assimilation}\label{alg:pdmda}
    \noindent\rule{\linewidth}{0.8pt}
    \begin{algorithmic}[1]
        \State Generate a training dataset of simulations with  $\nbparams$ parameter points $\theta_i, i=1,...,\nbparams$.
        \State Compute a POD basis $\Phi$ from the training manifold.
        \State Compute the DMD from the experimental measurements  to obtain $\Psi^{meas}$ and $\Lambda^{meas}$.
        \State Set $\crossg=\riT\Phi$ and solve $\crossg^T\crossg\tilde{\textbf{V}}_p = \crossg^T\Psi^{meas}$ for $\tilde{\textbf{V}}_p$.
        \State Compute the full-state modes $\Psi = \Phi \tilde{\textbf{V}}_p$. 
        \State Estimate reconstruction at $t=0$: $\textbf{c}= (\crossg^T\crossg)^{-1} \crossg^T \bu_0$.
        \State Compute initial DMD amplitudes $\textbf{b} = \Psi^{\dagger} \Phi \textbf{c}$.
        \State Reconstruct the full-state evolution $x(t)=\sum_{k=1}^r\psi_kb_ke^{\omega_kt}$
    \end{algorithmic}
    \noindent\rule{\linewidth}{0.8pt}
\end{algorithm}

The presented algorithm can be seen as a generalization of Compressed-sensing DMD \cite{brunton2015jcd} that takes into account parametric effects. Compressed-sensing DMD can be used to recover full dynamic states from sparse measurements assuming that the measurements can be represented in a sparse basis. This leads to an optimization problem that seeks for the sparsest mode representation consistent with the low-dimensional data, without considering parametric effects, as RODAS does.

\section{Numerical examples}
\label{sec:numerical_examples}

We benchmark RODAS in two scenarios: we first assess the capability of the assimilation strategy to recover a flow past a cylinder where the parametrization only relies on the Reynolds number. The goal of this preliminary study is to present a case where we can identify a nearest neighbor in the solution manifold for a given set of measurements, only looking at the dominant frequency on the DMD spectrum. This way, one could set a simple yet eloquent benchmark to test our method against. 

Second, we build a more challenging scenario, where the parameter identification with a single frequency is not feasible due to a more complex physical configuration. In the second example, we test our method in a non-Newtonian flow, whose rheological parameters are included in the training manifold.




\subsection{Flow past a cylinder}

\begin{figure*}
    \includegraphics[width=0.75\textwidth]{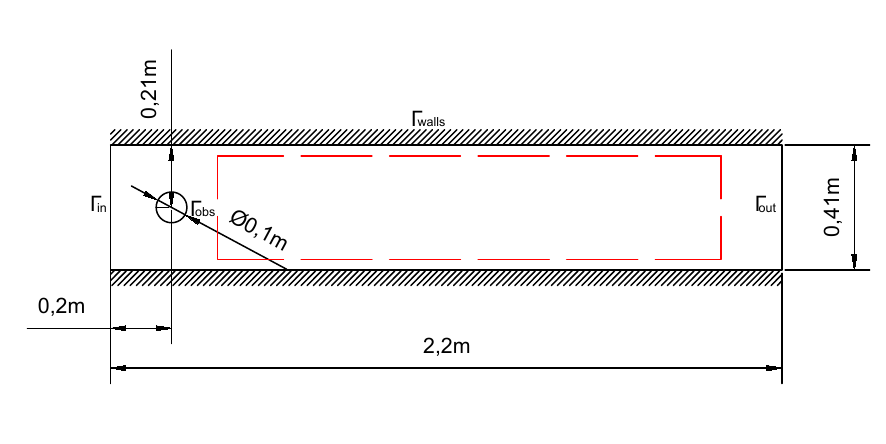}
    \caption{\label{im:vks_geom}Working domain for flow past a cylinder numerical example. The observation region is shown with a red dashed line.}
\end{figure*}

We consider a two-dimensional flow past a circular obstacle in the domain shown in Fig. \ref{im:vks_geom}. The geometric boundaries are split into disjoint sets as follows: $\partial \Omega = \Gamma_{\text{in}} \cup \Gamma_{\text{out}} \cup \Gamma_{\text{obs}} \cup \Gamma_{\text{walls}}$. Concerning the boundary conditions, we set a parabolic inflow on $\Gamma_{\text{in}}$, a no-slip boundary condition on $\Gamma_{\text{obs}} \cup \Gamma_{\text{walls}}$, and a traction-free condition on $\Gamma_{\text{out}}$. 

We compute 72 sample simulations of the solution manifold using the finite element method \cite{EG2013, volker2016} with the Multi-physics simulAtions for engineering and Data assimilation (MAD) software \cite{galarcemad, galarceThesis}, built on top of the linear algebra library PETSc \cite{petsc-web-page}. To solve the incompressible Navier-Stokes equations \eqref{eq:momentum}, \eqref{eq:mass}, we consider a lowest-order approximation of the finite element spaces, using a standard SUPG-PSPG stabilization approach. Additional details on the numerical aspects concerning the solution of the Navier-Stokes equations are given in Appendix A.


The fluid has water properties, with density $\rho=1$ kg/m$^3$ and dynamic viscosity $\mu=0.001$ Pa s. To sample the solution manifold, we draw a uniform distribution for the inflow mean velocity, keeping the Reynolds number in the range $100 \leq \text{Re} \leq 200$.
    
We use 60 simulations to build a parametric POD basis while the remaining 12 simulations were reserved for testing. Synthetic measurements were generated for each simulation based on the marked area depicted in Fig. \ref{im:vks_geom}, sampled at 280 grid-oriented measurement pixels.


\begin{figure*}[!htbp]
    \centering
    \includegraphics[width=0.7\textwidth]{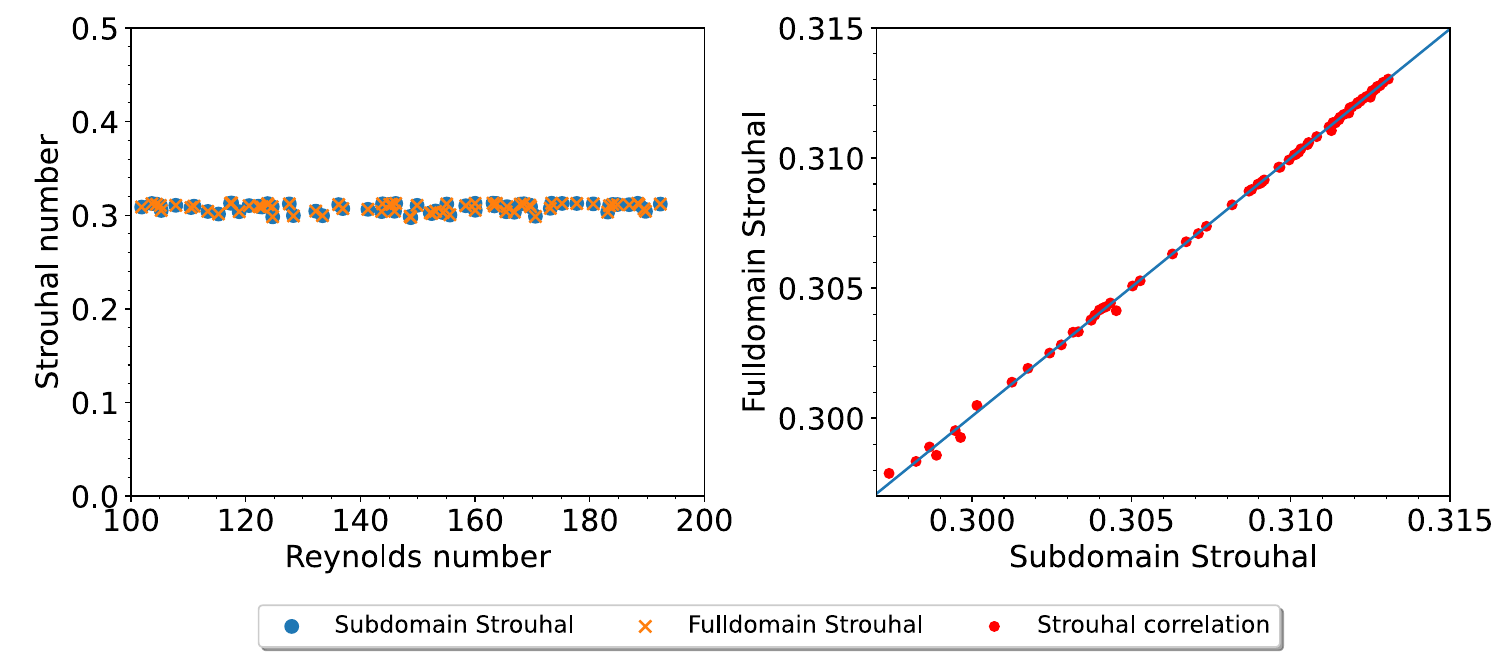}
    \caption{Validation of the dominant frequency extracted from localized measurements. Left: Strouhal number as a function of Reynolds number computed from the full domain and the observation subdomain, showing nearly identical estimates across the parameter range. Right: Correlation between Strouhal numbers obtained from the observation subdomain and the full domain.}
    \label{im:merged}
\end{figure*}

This first verification benchmark is set to understand if RODAS can outperform the naive approach of assigning to the measurements a single point in the solution manifold based purely on the dominant frequency. We can do this only in this overly simplified test case because the parameter space is essentially one dimensional. In any other more complex space, only RODAS can provide full state approximations. Thus, we compare:

\begin{enumerate}
    \item RODAS: reconstruction of the full domain state following algorithm described in section \ref{sec:rodas}.
    \item Dominant frequency recognition: reconstruction based solely on dominant frequency. We assign a complete simulation in the training set to a measured dataset with a nearest-neighbor approach.
\end{enumerate}

For the second approach, the frequency of the vortex shedding was computed using data-driven DMD only. For each simulation in the subdomain and full domain, we compute the frequency $f$ of each mode from the eigenvalues:
    \begin{equation}
        f_k=\frac{\text{imag}\{ \lambda_k \}}{2\pi\Delta t}.
    \end{equation}
In the previously mentioned range of Reynolds numbers, all the simulations present vortex shedding at different characteristic frequencies. For each simulation, the dominant vortex-shedding frequency was identified as the positive-frequency mode with the largest amplitude. Fig. \ref{im:merged} (left) shows good agreement between the ground-truth vortex shedding frequency and the dominant frequency extracted via DMD, and Fig. \ref{im:merged} (right) shows the estimated frequencies correlation between sub and full domain. 

\begin{figure}[!htbp]
    \centering
    \includegraphics[width=0.6\columnwidth]{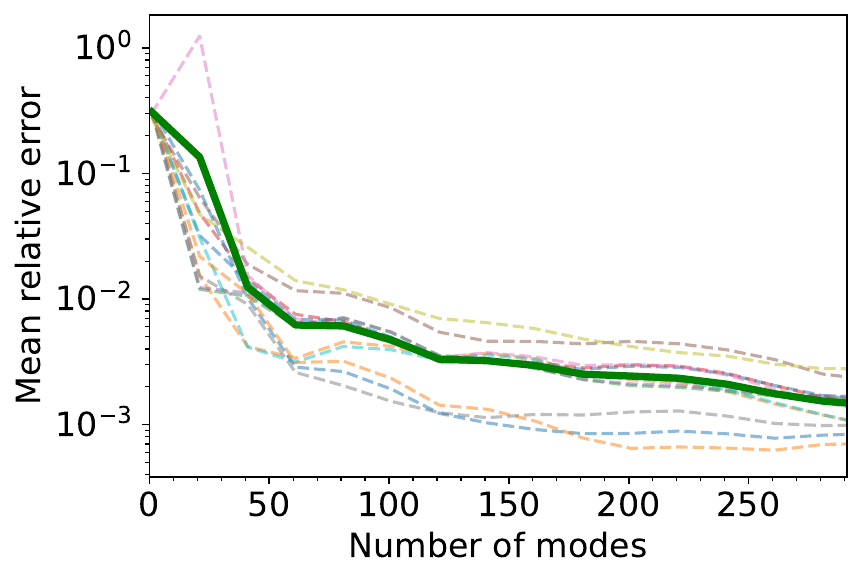}
    \caption{Mean relative error vs number of modes, for 12 test cases.}
    \label{im:en}
\end{figure}

    \def\nbsims{N_{\text{training}}}
    \def\nbrecsnaps{N_{\text{rs}}}
    
    
Next we study the reconstruction error obtained with RODAS when varying the number of spatial modes. For the $k-$th simulation, we compute the relative reconstruction error $E_k$ over $\nbrecsnaps$ time snapshots, and the mean error $E_t$ over all the test cases:
    \begin{equation}
        E_k=\frac{1}{\nbrecsnaps}\sum_{i=1}^{\nbrecsnaps} \frac{\lVert u_{i,k}^{gt}-u_{i,k}^{*} \rVert_2}{\lVert u_{i,k}^{gt}\rVert_2} \quad \text{and}\quad E_t=\sum_{k=1}^{\nbsims} \frac{E_k}{\nbsims}\, ,
        \label{eq:err}
    \end{equation}
where $u^{*}_{i,k}$ stands for the RODAS reconstruction of simulation $k$ at time instant $t_i$, while $u^{gt}_{i,k}$ is the corresponding ground truth state, $\nbsims=12$ the size of the test set, and $\nbrecsnaps$ the number of reconstructed snapshots. Although RODAS allows time super-resolution, we evaluate the error of the reconstruction in the time instants where the experimental measurements are provided.

    \begin{figure}[!htbp]
        \centering
        \includegraphics[width=0.6\columnwidth]{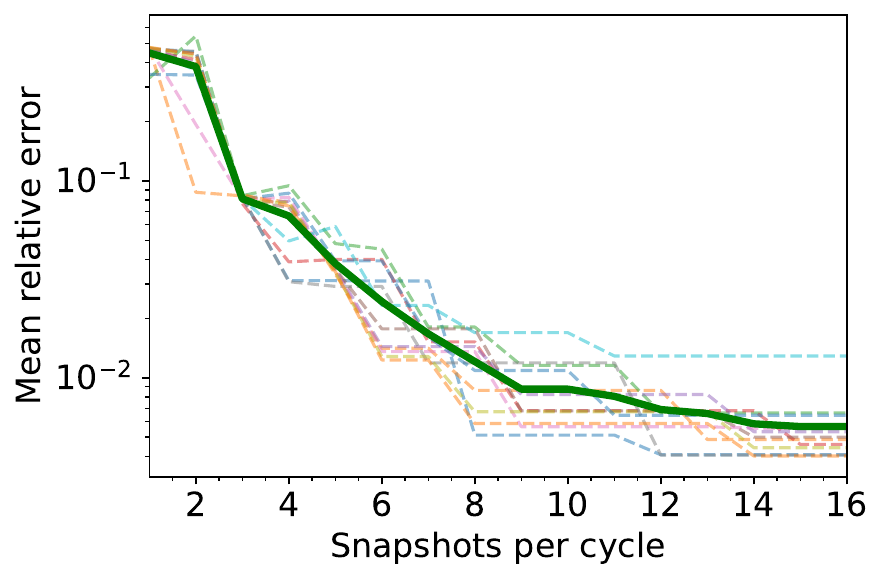}
        \caption{Time sensitivity related to the mean relative error of the reconstruction. Dashed lines correspond to particular test cases ($E_k$) and the continuous green line correspond to the mean of all test cases ($E_t$).}
        \label{im:timeSens}
    \end{figure}

    \begin{figure}[!htbp]
        \centering
        \includegraphics[width=0.8\columnwidth]{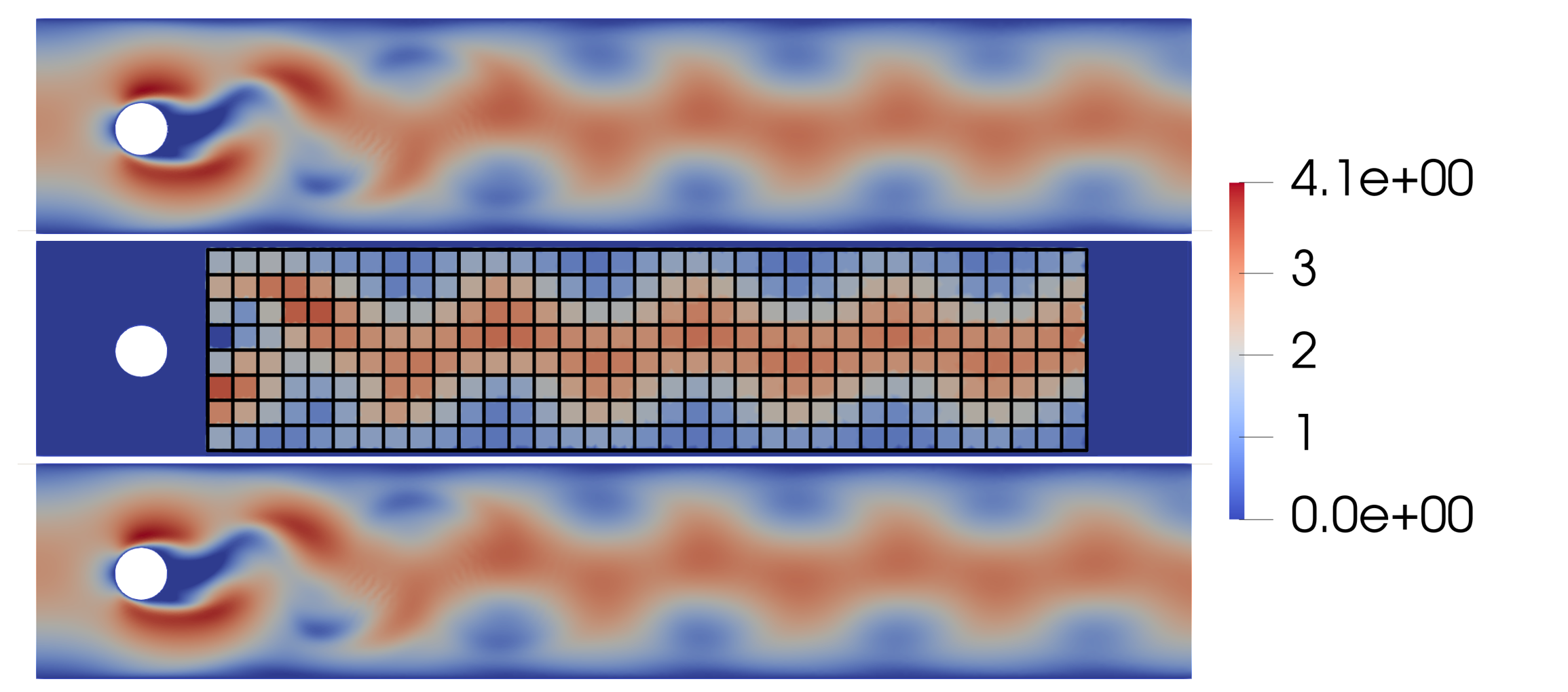}
        \caption{Comparison between the ground truth (top), measurements (middle) and reconstruction (bottom) for the velocity field in $x$ direction.}
        \label{im:vks_meas}
    \end{figure}
    
Fig. \ref{im:en} shows the mean relative error of reconstructed states using RODAS basis varying the number of POD modes in the parametric basis $\Phi$ in \eqref{eq:parametricPOD}. We observe that by using 17 measurements per cycle and 45 modes, the relative error drops below the $1 \%$, while if the dimension of the ROM increases further, the error approaches the $0.1 \%$. This compares with the approximation error of using the frequency nearest-neighbor. While still accurate in this admittedly simple scenario, the approach shows an average approximation error of $0.86 \%$.

\begin{table}
    \caption{Mean drag coefficient results for each test case}
    \label{tab:drags}
    \begin{center}
        \begin{tabular}{cccc}
        $Re$ & $\bar{C}_D^{gt}$ & $\bar{C}_D^{rec}$ & Relative error (\%) \\ \hline
        120.601 & 2.5497 & 2.5458 & 0.15296 \\
        123.895 & 2.5455 & 2.5546 & 0.35749 \\
        133.411 & 2.5194 & 2.5546 & 1.39716 \\
        152.195 & 2.4870 & 2.5005 & 0.54282 \\
        160.040 & 2.4716 & 2.4634 & 0.33177 \\
        163.361 & 2.4665 & 2.4729 & 0.25948 \\
        163.715 & 2.4654 & 2.4590 & 0.25959 \\
        165.575 & 2.4633 & 2.4653 & 0.08119 \\
        168.510 & 2.4602 & 2.4692 & 0.36582 \\
        173.123 & 2.4541 & 2.4524 & 0.06927 \\
        184.017 & 2.4397 & 2.4393 & 0.01640 \\
        188.369 & 2.4331 & 2.4189 & 0.58362 \\
        \end{tabular}
    \end{center}
\end{table}

\begin{figure}[!htbp]
        \centering
        \includegraphics[width=0.6\columnwidth]{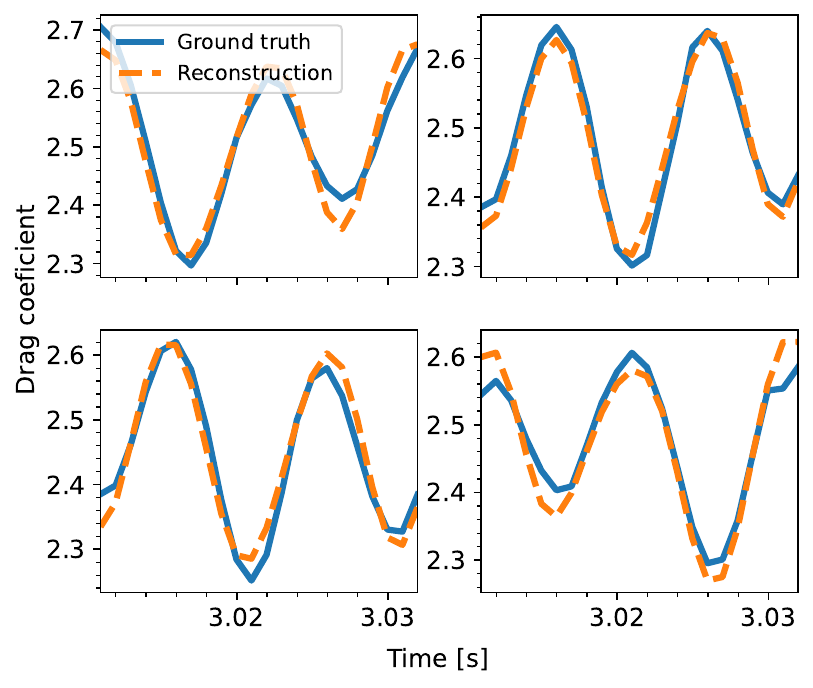}
        \caption{Comparison between the ground truth and the reconstruction drag coefficient curves in the first cases of Tab.~\ref{tab:drags}, considering a low sampling rate of 5 snapshots per cycle and 45 modes.}
        \label{im:dragcomp}
\end{figure}


We next assess the sensitivity of the method to temporal under-sampling by varying the number of snapshots per vortex-shedding cycle. Similar to the previous case, we generate reconstructions for each test case and compute the mean relative error. Fig. \ref{im:timeSens} shows the results, where the mean relative error goes below $3\%$ when using 5 snapshots per cycle, which represents one of the worst possible sampling scenarios, but yields a tolerable error for most applications.

Finally, we generate full-state reconstructions using the RODAS and synthetic measurements following Tab.~\ref{alg:pdmda}, and using 45 modes and a sampling rate of 5 snapshots per cycle. Fig. \ref{im:vks_meas} shows an example of the reconstruction and the resolution of the generated measurements. As an additional validation, Fig. \ref{im:dragcomp} compares the drag coefficient computed from the reconstructed fields against the ground truth, and Tab.~\ref{tab:drags} summarizes the results for each test case. We convey a variational strategy to compute drag forces, described in detail in Appendix B.

\label{subsec:vdrag}
\def\velfield{\textbf{u}}
\def\pressfield{\textbf{p}}
\def\testmom{\textbf{v}}
\def\testcont{\textbf{q}}
\def\matmass{\textbf{M}}
\def\matrig{\textbf{K}}
\def\matb{\textbf{B}}
\def\matconv{\textbf{C}}
\def\normvec{\hat{\textbf{n}}}
\def\bound{S}
\def\nbnodes{NN}


Most of the test cases present errors below $1 \%$ for the mean of the drag coefficient, and Fig. \ref{im:dragcomp} shows that the method is capable of capturing both amplitude and frequency. 

\def\sigpower{P_s}
\def\noisepower{P_n}
\def\noise{\textbf{X}_n}
\def\snapmatnoise{\textbf{Y}}
\def\snapmatmeas{\textbf{X}}

\subsection{Non-Newtonian von Kármán vortex street}

For this example we consider the same geometry from Fig. \ref{im:vks_geom}, with the same boundary conditions. What differs is that we now test the ability of RODAS to capture the rheology effect in a more complex scenario. We used three test cases per rheology, considering pseudo-plastic, Newtonian and shear-thickening fluids in the same parametric manifold, i.e., assuming a broad rheology uncertainty in the prediction task.

    \begin{figure}[htbp]
        \centering
        \includegraphics[width=0.6\columnwidth]{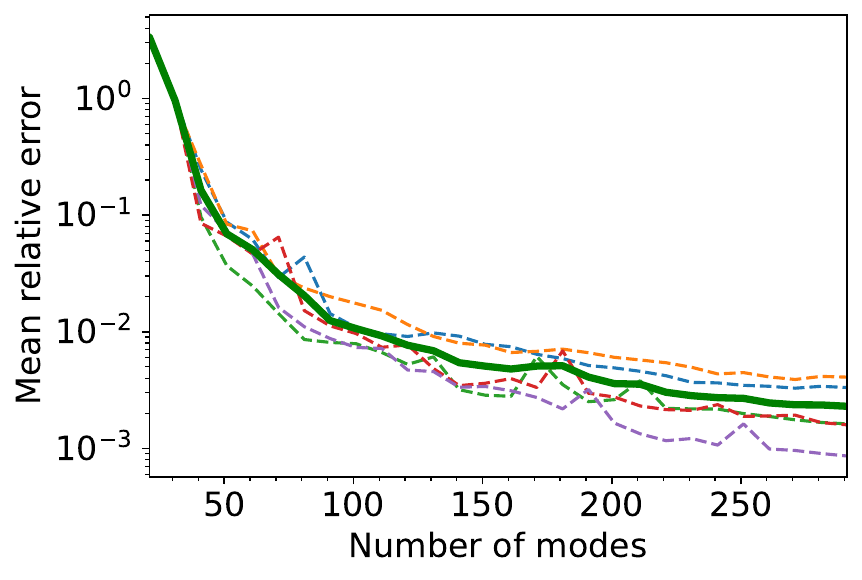}
        \caption{Number of DMD modes versus mean relative error according to Eq.~(\ref{eq:err}), from 21 to 291 modes.}
        \label{im:vksnn_nbmodes}
    \end{figure}

For the training manifold 192 simulations were sampled, varying the inlet velocity and the power-law index, which controls the rheology behavior. For this, we used the Carreau-Yasuda model, where the apparent viscosity in terms of the shear rate $\eta(\dot{\gamma})$ follows the formula

    \begin{equation}
        \eta(\dot{\gamma})=\eta_{\infty} + (\eta_0-\eta_{\infty})[\lambda \dot{\gamma}^a]^{\frac{n-1}{a}} \;,
    \end{equation}
    where $a$ is the Yasuda parameter, which controls the transition sharpness, $n$ is the power-law index, $\mu_0$ is the low-shear viscosity, $\mu_{\infty}$ the high-shear viscosity, $\lambda$ is a time constant.

The simulations considered $\mu_0=0.005$ Pa $\cdot$ s, $\mu_{\infty}= 0.001$ Pa $\cdot$ s, $a=1.25$ and $\lambda=0.02$ s. The training manifold considers power-law index values of $0.75$, $1.0$ and $1.25$, and Reynolds numbers at low-shear between 100 and 200, computed as:
    \begin{equation}
        Re_0 = \frac{\rho \bar{u} L}{\eta_0}\;,
    \end{equation}
drawn from uniform distributions in the aforementioned ranges.

Concerning the parametric ROM dimension, multiple reconstructions were generated using synthetically measured data with 10 snapshots per cycle. As expected, all test reconstructions required more modes to have similar errors as the Newtonian case. We choose to use 150 modes for the following reconstructions, as the mean relative error drops near $1\%$ in all cases, according to Fig. \ref{im:vksnn_nbmodes}.
    

\begin{figure*}
    \includegraphics[width=0.8\textwidth]{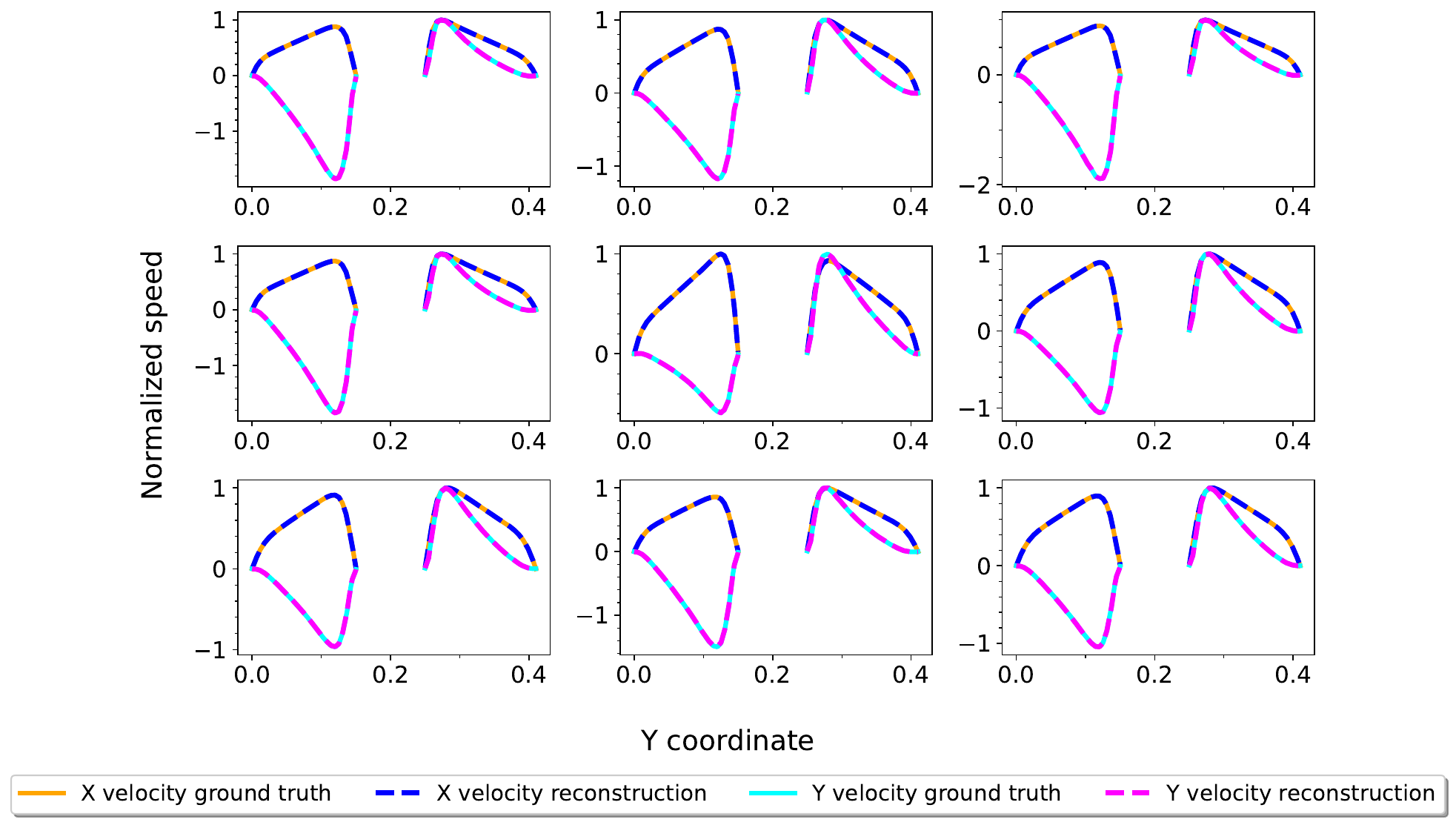}
    \caption{\label{im:vksnn_rheo}Velocity profile reconstructions for pseudo-plastic (first row), Newtonian (second row), and thickening (third row)}
\end{figure*}
    
Despite the low relative error in the reconstructions, we test the ability to reconstruct the rheology effects by comparing the velocity profile at the sides of the obstacle. Fig. \ref{im:vksnn_rheo} compares ground truth and reconstruction velocity components along a line perpendicular to the flow direction, and shows a good agreement in all cases. We remark the flexibility of RODAS to identify the right basis that matches different measurement scenarios, in spite of the broad rheological range of the parametric manifold.

\section{Conclusions}

We have proposed, discussed and tested a novel data assimilation procedure to deal with low spatio-temporal resolution data. Our numerical findings support our idea that RODAS can estimate a full state trajectory approximating a dynamical system by merging a low-resolution DMD reconstruction from the measurement data, and a parametric high-fidelity POD from numerical simulations, which is, to the best of the authors’ knowledge, an original technique involving recent manifold learning concepts. 

We remark the efficiency of the method for estimating high-resolution fields from localized and eventually poor data. RODAS is able not only to do time super-resolution (which is in fact a feature inherited from DMD), but also to extrapolate fields beyond experimental acquisition regions. This, along with the fact that RODAS delivers real-time estimates, is the main strength of our approach. 

We pose as further research the following three aspects: 1) the integration of real laboratory data for a controlled yet more realistic testing scenario, 2) a mathematical root of the method in terms of an infinite dimensional formulation analysis, including error bounds and uncertainty quantification, and 3) the incorporation of recently introduced modifications of DMD, such as piDMD.

\appendix
\section{Forward simulations}

We describe next the approach to solve numerically the Navier-Stokes equations using the finite element method, including the weak form, approximation spaces, and stabilization strategies.

We start from the incompressible Navier-Stokes equations, with no external forces:
\begin{equation}
    \begin{aligned}
    \frac{\partial}{\partial t} \velfield + (\velfield\cdot\nabla)\velfield - \nabla\cdot(2\eta(\velfield)\nabla^s \velfield) + \nabla\pressfield &=0\;, \\
    \nabla\cdot\velfield &=0\;,
    \end{aligned}
\end{equation}
where $\velfield$ is the velocity field, $\pressfield$ is the pressure field, $\eta(\velfield)$ is the apparent viscosity and $\nabla^s$ is the symmetric gradient.

We solve for the state functions $\velfield$ and $\pressfield$ using the finite element method \cite{EG2013}, so we derive next the weak form of the governing equations:
\begin{equation}
\begin{aligned}
    ( \frac{\partial \velfield}{\partial t}, \testmom )
    + 
    ( (\velfield \cdot \nabla)\velfield, \testmom ) 
    +
    ( 2\eta \nabla^s \velfield, \eta \nabla^s \testmom )
    - ( \pressfield,\nabla\cdot v ) \\
    + (\nabla \cdot \velfield, q)- ( (2\eta \nabla^s \velfield - p \textbf{I})\normvec, v )_{\partial \Omega} 
    =0, \\  \forall (v, ~q) \in H^1(\Omega) \times L^2(\Omega) \;.
   \end{aligned} 
   \label{eq:ns_weak}
\end{equation}



The time derivative is approximated by using an optimized second-order backward differentiation formula (BDF2). In addition, to deal with the nonlinear nature of the equations, we apply a Picard scheme plus the Anderson acceleration strategy \cite{barnafi2026}, with a memory size of three iterated solutions.

Concerning the space discretization, we use piecewise linear basis functions for both velocity and pressure. This equal-order approach  violates inf-sup stability \cite{volker2016}. Thus, we apply a Streamline Upwind/Petrov-Galerkin stabilization (SUPG), and Pressure Stabilized Petrov–Galerkin stabilization (PSPG) terms in order to reduce oscillations along streamlines and circumvent the incompressibility constraint. We therefore compute time-marching solutions for the weak form \eqref{eq:ns_weak} adding the following stabilization terms:
$$
\begin{aligned}
R(u,v,p,q) + \tau_{supg} (u^n \cdot \nabla u^{n+1,k},u^n\cdot\nabla v)
+\tau_{supg} (\nabla \pressfield, u^n\cdot \nabla v) \\
+\tau_{supg} (BDF2^{opt}, u^n \cdot \nabla v)
+\tau_{pspg}(\nabla \pressfield, \nabla q) \\
+\tau_{pspg} (u^n\cdot\nabla u^{n+1,k} , \nabla q)
+\tau_{pspg} (BDF2^{opt} , \nabla q), 
\\  \forall (v, ~q) \in H^1(\Omega) \times L^2(\Omega), \;.
\end{aligned}
$$
where $BDF2^{opt}$ stands for the backward Euler approximation of the time derivatives, $R(u,p,v,q)$ stands for the momentum and mass weak residuals from \eqref{eq:ns_weak}, and the stabilization parameters are calculated as: 
$$
\tau_{supg}=\tau_{supg}=\tau_K
=
\left[
\left(\frac{2}{\Delta t}\right)^2
+
\left(\frac{2|\mathbf{u}|}{h_K}\right)^2
+
C_I^2
\left(\frac{\eta_K}{h_K^2}\right)^2
\right]^{-1/2},
$$
where the mesh size $h_K$ and apparent viscosity $\eta_K$ are computed element-wise.

\section{Variational drag estimation}

In this appendix we describe the method used for computing the drag force based on the variational form of the momentum equation, taking inspiration from \cite{volker2016_femFlows}. The drag force acting on a body immersed in an incompressible fluid depends on the pressure and velocity gradient fields. It is defined as:
\begin{equation}
    \label{eq:drag}
    \textbf{F}_D = \left( \int_S (-\pressfield\textbf{I}+\tau)\cdot\normvec ~\text{dS} \right) \cdot \hat{w} \;,
\end{equation}
where $\pressfield$ is the pressure field, $\normvec$ is the unit outward normal vector on the surface $S$, $\hat{w}$ is the unit vector in the drag direction, and $\tau$ is the viscous stress tensor, dependent on the velocity field and related to the rheology model.

When reduced-order or data-driven reconstructions are used, the gradients of the learned fields may not be sufficiently accurate for directly applying Eq.~(\ref{eq:drag}). Furthermore, pressure data may be unavailable if the dataset originates from velocimetry methods. To address this, the drag force can be computed through the variational form of the momentum equation. Consider the Navier-Stokes momentum equation, with negligible external forces for incompressible flows:
\begin{equation}
    \label{eq:mom}
    \rho \frac{\partial \velfield}{\partial t}+ \rho \velfield \cdot\nabla\velfield  -\nabla \cdot \tau +\nabla \pressfield = \textbf{0} \; .
\end{equation}

Following the same functional setting of Appendix A, a standard weak form of these equations look like:
\begin{equation}
    \begin{aligned}
    \label{eq:weak-1}
    \rho \left(   \int_{\Omega}{\frac{\partial \velfield}{\partial t} \cdot\testmom\; d\Omega}  +  \int_{\Omega}{(\velfield\cdot\nabla\velfield) \cdot\testmom\; d\Omega}   \right) \\ -\int_{\Omega}{\tau \cdot\testmom\; d\Omega}  +  \int_{\Omega}{\nabla \pressfield \cdot\testmom\; d\Omega} = \textbf{0}, \quad \forall \testmom \in H^1(\Omega).
    \end{aligned}
\end{equation}

Integrating by parts leads to:
\begin{equation}
    \label{eq:ens_var}
    \begin{aligned}
        \rho\left(   (\frac{\partial\velfield}{\partial t},\testmom) _\Omega 
        + ( \velfield \cdot\nabla\velfield,\testmom) _\Omega  \right) + ( \tau,\nabla\testmom)_{\Omega} \\ - ( \tau \hat{\textbf{n}},\testmom)\ _{\partial\Omega} + ( \nabla \cdot \testmom, \pressfield ) _{\Omega}+ ( \pressfield\normvec, \testmom )_{\partial\Omega} = \textbf{0}\\
        \forall\; \testmom \in H^{1}(\Omega)\;,
        \end{aligned}
\end{equation}


Next we can choose a convenient test function to decompose the load vector in the drag force direction. To compute this function we solve an auxiliary problem on the flow domain with convenient boundary conditions. If we consider the setup shown in Fig. \ref{im:vks_geom}, we can solve a Stokes problem with equations:
$$
\begin{aligned}
    - \mu\Delta\testmom + \nabla\pressfield &= \textbf{0}, \\
    \nabla\cdot \testmom &= \textbf{0}.
\end{aligned}
$$
And set $\velfield=(0,0)$ on $\Gamma_{in} \cup \Gamma_{walls}$ and $\Gamma_{obs}=(1,0)$, so we can solve for the test function $\testmom$. This way we impose unit vectors at the obstacle boundary in the drag direction, and obtain a $H^1$ divergence-free velocity field, allowing us to use it as a test function and ignore the pressure term.

Finally, the drag force is obtained by inserting the computed test function and discrete solutions for velocity and pressure in Eq.~(\ref{eq:ens_var}). 

\bibliographystyle{plainnat}
\bibliography{references}

\end{document}